\documentclass[aps,preprint,letterpaper,balancelastpage ,superscriptaddress,showpacs,prd]{revtex4}
\usepackage{amsfonts}
\usepackage{amsmath}
\usepackage{amssymb}
\usepackage{eurosym}
\usepackage{paralist}
\usepackage{graphicx}
\usepackage[colorlinks=true,linkcolor=blue,citecolor=blue, filecolor=blue,urlcolor=blue]{hyperref}

\begin{document}

\title{Static solutions in Einstein-Chern-Simons gravity}
\author{J. Cris\'ostomo}
\email{jcrisostomo@udec.cl}
\affiliation{Departamento de F\'{\i}sica, Universidad de Concepci\'{o}n, Casilla 160-C,
Concepci\'{o}n, Chile}
\author{F. Gomez}
\email{fernagomez@udec.cl}
\affiliation{Departamento de F\'{\i}sica, Universidad de Concepci\'{o}n, Casilla 160-C,
Concepci\'{o}n, Chile}
\author{P. Mella}
\email{patriciomella@udec.cl}
\affiliation{Departamento de F\'{\i}sica, Universidad de Concepci\'{o}n, Casilla 160-C,
Concepci\'{o}n, Chile}
\author{C. Quinzacara}
\email{cristian.cortesq@uss.cl}
\affiliation{Departamento de F\'{\i}sica, Universidad de Concepci\'{o}n, Casilla 160-C,
Concepci\'{o}n, Chile}
\affiliation{Facultad de Ingenier\'ia y Tecnolog\'ia, Universidad San Sebasti\'an, Campus
Las Tres Pascualas, Lientur 1457, Concepci\'{o}n, Chile}
\author{P. Salgado}
\email{pasalgad@udec.cl}
\affiliation{Departamento de F\'{\i}sica, Universidad de Concepci\'{o}n, Casilla 160-C,
Concepci\'{o}n, Chile}
\date{\today }

\begin{abstract}
In this paper we study static solutions with more general symmetries than
the spherical symmetry of the five-dimensional Einstein-Chern-Simons
gravity. In this context, we study the coupling of the extra bosonic field $%
h^{a}$ with ordinary matter which is quantified by the introduction of an
energy-momentum tensor field associated with $h^{a}$. It is found that exist
\begin{inparaenum}[(i)]
	\item a negative tangential pressure zone around low-mass
	distributions ($\mu <\mu _{1}$) when the coupling constant $\alpha $ is
	greater than zero; 
	\item a maximum in the tangential pressure, which can be
	observed in the outer region of a field distribution that satisfies $\mu
	<\mu _{2}$; 
	\item solutions that behave like those obtained from models
	with negative cosmological constant. 
\end{inparaenum} In such a situation, the field $h^{a}$
plays the role of a cosmological constant.
\end{abstract}

\pacs{04.70.Bw, 11.15.Yc, 04.90.+e, 04.50.Gh}
\maketitle

\section{Introduction}

Five-dimensional Einstein-Chern-Simons gravity (EChS) is a gauge theory
whose Lagrangian density is given by a 5-dimensional Chern-Simons form for
the so called $\mathfrak{B}$ algebra \cite{salg1}. This algebra can be
obtained from the AdS algebra and a particular semigroup $S$ by means of the 
$S$-expansion procedure introduced in Refs. \cite{salg2,salg3}. The field content induced by the $\mathfrak{B}$ algebra includes the
vielbein $e^{a}$, the spin connection $\omega ^{ab}$, and two extra bosonic
fields $h^{a}$ and $k^{ab}.$ The EChS gravity has the interesting property
that the five dimensional Chern-Simons Lagrangian for the $\mathfrak{B}$
algebra, given by \cite{salg1}:

\begin{equation}
L_{\mathrm{EChS}}=\alpha _{1}l^{2}\varepsilon
_{abcde}R^{ab}R^{cd}e^{e}+\alpha _{3}\varepsilon _{abcde}\left( \frac{2}{3}%
R^{ab}e^{c}e^{d}e^{e}+2l^{2}k^{ab}R^{cd}T^{\text{ }e}+l^{2}R^{ab}R^{cd}h^{e}%
\right) ,  \label{1}
\end{equation}%
where $R^{ab}=\mathrm{d}\omega ^{ab}+\omega _{\text{ }c}^{a}\omega ^{cb}$
and $T^{a}=\mathrm{d}e^{a}+\omega _{\text{ }c}^{a}e^{c}$, leads to the
standard General Relativity without cosmological constant in the limit where
the coupling constant $l$ tends to zero while keeping the effective Newton's
constant fixed \cite{salg1}.

In Ref. \cite{salg4} was found a spherically symmetric solution for the
Einstein-Chern-Simons field equations and then was shown that the standard
five dimensional solution of the Einstein-Cartan field equations can be
obtained, in a certain limit, from the spherically symmetric solution of
EChS field equations. The conditions under which these equations admit black
hole type solutions were also found.

The purpose of this work is to find static solutions with more general
symmetries than the spherical symmetry. These solutions are represented by
three-dimensional maximally symmetric spaces: open, flat and closed.

The functional derivative of the matter Lagrangian with respect to the field 
$\ h^{a}$ is considered as another source of gravitational field, so that it
can be interpreted as a second energy-momentum tensor: the energy-momentum
tensor for field $\ h^{a}$. This tensor is modeled as an anisotropic fluid,
the energy density, the radial pressure and shear pressures are
characterized. The results lead to identify the field $h^{a}$ with the
presence of a cosmological constant. The spherically symmetric solutions of
Ref. \cite{salg4} can be recovered from the general static solutions.

The article is organized as follows: In section \ref{sectionII} we briefly
review the Einstein-Chern-Simons field equations together with their
spherically symmetric solution, which lead, in certain limit, to the
standard five-dimensional solution of the Einstein-Cartan field equations.
In section \ref{sectionIII} we obtain general static solutions for \ the
Einstein-Chern-Simons field equations. The obtaining of the energy momentum
tensor for the field $h^{a},$ together with the conditions that must be
satisfied by the energy density and radial and tangential pressures, also
will be considered in section \ref{sectionIII}. In section \ref{sectionIV}
we recover the spherically symmetric black hole solution found in Ref. \cite{salg4} from the general static solutions and will study the energy density
and radial and tangential pressures for a naked singularity and black hole
solutions. Finally, concluding remarks are presented in section \ref{sectionV}.

\section{Spherically symmetric solution of \textbf{EChS field equations } 
\label{sectionII}}

In this section we briefly review the Einstein-Chern-Simons field equations
together with their spherically symmetric solution. We consider the field
equations for the Lagrangian 
\begin{equation}
L=L_{\mathrm{EChS}}+L_{\mathrm{M}},  \label{accion01}
\end{equation}%
where $L_{\mathrm{EChS}}$ is the Einstein-Chern-Simons gravity Lagrangian
given in (\ref{1}) and $L_{\mathrm{M}}$ is the corresponding matter
Lagrangian.

In the presence of matter described by the langragian $L_{\mathrm{M}}=L_{%
\mathrm{M}}(e^{a},h^{a},\omega ^{ab}),$ the field equations obtained from
the action (\ref{accion01}) when $T^{a}=0$ and $k^{ab}=0$ are given by \cite{salg4}:

\begin{align}
& \mathrm{d}e^{a}+\omega _{\text{ }b}^{a}e^{b}=0,  \label{3.1} \\
& \varepsilon _{abcde}R^{cd}\mathrm{D}_{\omega }h^{e}=0,  \label{3.2} \\
& \alpha _{3}l^{2}\star \left( \varepsilon _{abcde}R^{bc}R^{de}\right)
=-\star \left( \frac{\delta L_{\mathrm{M}}}{\delta h^{a}}\right) ,
\label{3.3} \\
& \star \left( \varepsilon _{abcde}R^{bc}e^{d}e^{e}\right) +\frac{1}{2\alpha 
}l^{2}\star \left( \varepsilon _{abcde}R^{bc}R^{de}\right) =\kappa _{\mathrm{%
EH}}\hat{T}_{a},  \label{3.4}
\end{align}%
where $\mathrm{D}_{\omega }$ denotes the exterior covariant derivative
respect to the spin connection $\omega $, \textquotedblleft $\star $%
\textquotedblright\ is the Hodge star operator, $\alpha :=\alpha _{3}/\alpha
_{1}$, $\kappa _{\mathrm{EH}}$ is the coupling constant in five-dimensional
Einstein-Hilbert gravity, 
\begin{equation*}
\hat{T}_{a}=\hat{T}_{ab}e^{b}=-\star \left( \frac{\delta L_{\mathrm{M}}}{%
\delta e^{a}}\right)
\end{equation*}%
is the energy-momentum 1-form, with $\hat{T}_{ab}$ the usual energy-momentum
tensor of matter fields, and where we have considered, for simplicity, $%
\delta L_{M}/\delta \omega ^{ab}=0.$

Since equation (\ref{3.4}) is the generalization of the Einstein field
equations, it is useful to rewrite it in the form%
\begin{equation}
\star \left( \varepsilon _{abcde}R^{bc}e^{d}e^{e}\right) =\kappa _{\mathrm{EH%
}}\hat{T}_{a}+\frac{1}{2\alpha \alpha _{3}}\star \left( \frac{\delta L_{%
\mathrm{M}}}{\delta h^{a}}\right)
\end{equation}%
where we have used the equation (\ref{3.3}). This result leads to the
definition of the 1-form energy-momentun associated with the field $h^{a}$%
\begin{equation}
\hat{T}_{a}^{(h)}=\hat{T}_{ab}^{(h)}e^{b}=\frac{1}{2\alpha \alpha _{3}}\star
\left( \frac{\delta L_{\mathrm{M}}}{\delta h^{a}}\right) .
\end{equation}%
This allows to rewrite the field equations (\ref{3.3}) and (\ref{3.4}) as

\begin{align}
&-\mathrm{sgn}(\alpha )\frac{1}{2}l^{2}\star \left( \varepsilon
_{abcde}R^{bc}R^{de}\right) =\kappa _\mathrm{EH}\hat{T}_{a}^{(h)},
\label{7.3} \\
&\star \left( \varepsilon _{abcde}R^{bc}e^{d}e^{e}\right) +\mathrm{sgn}%
(\alpha )\frac{1}{2}l^{2}\star \left( \varepsilon
_{abcde}R^{bc}R^{de}\right) =\kappa _\mathrm{EH}\hat{T}_{a},  \label{7.4}
\end{align}%
where the absolute value of the constant $\alpha $ has been absorbed by
redefining the parameter $l$%
\begin{equation*}
l\rightarrow l^{\prime }=\frac{1}{\sqrt{\left\vert \alpha \right\vert }}=%
\sqrt{\left\vert \frac{\alpha _{1}}{\alpha _{3}}\right\vert } .
\end{equation*}

\subsection{\textbf{Static and spherically symmetric solution}}

In this subsection we briefly review the spherically symmetric solution of
the EChS field equations, which lead, in certain limit, to the standard
five-dimensional solution of the Einstein-Cartan field equations.

In five dimensions the static and spherically symmetric metric is given by 
\begin{equation*}
\mathrm{d}s^{2}=-f^{2}(r)\mathrm{d}t^{2}+\frac{\mathrm{d}r^{2}}{g^{2}(r)}%
+r^{2}\mathrm{d}\Omega _{3}^{2}=\eta _{ab}e^{a}e^{b},
\end{equation*}%
where $\mathrm{d}\Omega _{3}^{2}=\mathrm{d}\theta _{1}^{2}+\sin ^{2}\theta
_{1}\mathrm{d}\theta _{2}^{2}+\sin ^{2}\theta _{1}\sin ^{2}\theta _{2}%
\mathrm{d}\theta _{3}^{2}$ is the line element of 3-sphere $S^{3}$.

%and $\eta
%_{ab}=\mathrm{diag}(-1,+1,+1,+1,+1)$ is the Minkowski metric.

Introducing an orthonormal basis 
\begin{align}
e^{T}=f(r)\,\mathrm{d}t,\ e^{R}=\frac{\mathrm{d}r}{g^2(r)},\ e^{1}=r\,%
\mathrm{d}\theta _{1},  \notag \\
e^{2}=r\sin \theta _{1}\,\mathrm{d}\theta _{2},\ e^{3}=r\sin \theta _{1}\sin
\theta _{2}\, \mathrm{d}\theta _{3}  \label{10}
\end{align}%
and replacing into equation (\ref{7.4}) in vacuum ($\hat{T}_{TT}=\hat{T}%
_{RR}=\hat{T}_{ii}=0$), we obtain the EChS field equations for a spherically
symmetric metric equivalent to eqs. ({\color{blue}26} - {\color{blue}28})
from Ref. \cite{salg4}.

\subsubsection{Exterior solution}

Following the usual procedure, we find the following solution \cite{salg4}: 
\begin{equation}
f^2(r)=g^2(r)=1+\text{sgn}(\alpha )\left(\frac{r^{2}}{l^{2}}-\beta \sqrt{%
\frac{r^{4}}{l^{4}}+\text{sgn}(\alpha )\frac{\kappa _\mathrm{EH}}{6\pi
^{2}l^{2}}M}\right),  \label{14}
\end{equation}%
where $M$ is a constant of integration and $\beta =\pm 1$ shows the
degeneration due to the quadratic character of the field equations. From (%
\ref{14}) it is straightforward to see that when $l\rightarrow 0$, it is
necessary to consider $\beta =1$ to obtain the standard solution of the
Einstein-Cartan field equation, which allows to identify the constant $M$,
with the mass of distribution.

\section{General Static Solutions with General Symmetries\label{sectionIII}}

In Ref. \cite{salg4} were studied static exterior solutions with spherically
symmetry for the Einstein-Chern-Simons field equations in vacuum. In this
reference were found the conditions under which the field equations admit
black holes type solutions and were studied the maximal extension and
conformal compactification of such solutions.

In this section we will show that the equations of Einstein-Chern-Simons
allow more general solutions that found for the case of spherical symmetry.
The spherical symmetry condition will be relaxed so as to allow studying
solutions in the case that the space-time is foliated by maximally symmetric
spaces more general than the 3-sphere. It will also be shown that, for
certain values of the free parameters, these solutions lead to the solutions
found in Ref. \cite{salg4}.

\subsection{Solutions to the EChS field equations}

Following Refs. \cite{Oliv01,Oliv02}, we consider a static metric of
the form 
\begin{equation}
\mathrm{d}s^{2}=-f^{2}(r)\,\mathrm{d}t^{2}+\frac{\mathrm{d}r^{2}}{g^{2}(r)}%
+r^{2}\mathrm{d}\Sigma _{3}^{2}.  \label{18}
\end{equation}%
where $\mathrm{d}\Sigma _{3}^{2}$ is the line element of a three-dimensional
Einstein manifold $\Sigma _{3}$, which is known as the \emph{base manifold} 
\cite{7}.

Introducing an ortonormal basis, we have 
\begin{equation*}
e^{T}=f(r)\ \mathrm{d}t,\quad e^{R}=\frac{\mathrm{d}r}{g(r)},\quad e^{m}=r%
\tilde{e}^{m},
\end{equation*}%
where $\tilde{e}^{m}$, with $m=\{1,2,3\}$, is the \textit{dreibein} of the
base manifold $\Sigma _{3}$.

From eq. (\ref{3.1}), it is possible to obtain the spin connection in terms
of the vielbein. From Cartan's second structural equation $R^{ab}=\mathrm{d}%
\omega ^{ab}+\omega _{\text{ }c}^{a}\omega ^{cb}$ we can calculate the
curvature matrix. The nonzero components are 
\begin{align}
R^{TR}& =-\left( \frac{f^{\prime \prime }}{f}g^{2}+\frac{f^{\prime }}{f}%
g^{\prime }g\right) e^{T}e^{R},\ R^{Tm}=-\frac{f^{\prime }}{f}g^{2}e^{T}%
\tilde{e}^{m},  \notag \\
R^{Rm}& =-g^{\prime }g\,e^{R}\tilde{e}^{m},\ R^{mn}=\tilde{R}^{mn}-g^{2}%
\tilde{e}^{m}\tilde{e}^{n},  \label{22}
\end{align}%
where $\tilde{R}^{mn}=\mathrm{d}\tilde{\omega}^{mn}+\tilde{\omega}_{\text{ \ 
}p}^{m}\tilde{\omega}^{pn}$ are the components of the curvature of the base
manifold. To define the curvature of the base manifold is necessary to
define the spin connection $\tilde{\omega}^{mn}$ of the base manifold. This
connection can be determined in terms of the dreibein $\tilde{e}^{m}$ using
the property that the total covariant derivative of the vielbein vanishes
identically, and the condition of zero torsion $\tilde{T}^{m}=0$.

Replacing the components of the curvature (\ref{22}) in the field equations (%
\ref{7.4}), for the case where $\hat{T}_{a}=0$ (vacuum), we obtain three
equations

\begin{equation}
B_{u}(r)\tilde{R}(\tilde{x})+6A_{u}(r)=0,\quad u=\{0,1,2\},  \label{23.4}
\end{equation}%
%
%\begin{align}
%B_{T}(r)\tilde{R}(\tilde{x})+6A_{T}(r)& =0,  \label{23.1} \\
%B_{R}(r)\tilde{R}(\tilde{x})+6A_{R}(r)& =0,  \label{23.2} \\
%B(r)\tilde{R}^{mn}(\tilde{x})+6A(r)\tilde{e}^{m}\tilde{e}^{n}& =0,
%\label{23.3}
%\end{align}%
where $\tilde{R}(\tilde{x})$ is the Ricci scalar of the base manifold and
the functions $A_{u}(r)$ and $B_{u}(r)$ are given by 
\begin{align}
A_{0}(r)& =-2r\left( g^{2}r^{2}\right) ^{\prime }+\text{sgn}(\alpha
)\,l^{2}r\left( g^{4}\right) ^{\prime },  \label{24.1} \\
B_{0}(r)& =2r\left( 2r-\text{sgn}(\alpha )\,l^{2}\left( g^{2}\right)
^{\prime }\right) ,  \label{24.2} \\
A_{1}(r)& =2r\left( -2rg^{2}-3\,\text{sgn}(\alpha )\,l^{2}r^{2}g^{2}\frac{%
f^{\prime }}{f}+2\,\text{sgn}(\alpha )\,l^{2}g^{4}\frac{f^{\prime }}{f}%
\right) ,  \label{24.3} \\
B_{1}(r)& =2r\left( 2r-2\,\text{sgn}(\alpha )\,l^{2}g^{2}\frac{f^{\prime }}{f%
}\right) ,  \label{24.4} \\
A_{2}(r)& =-2r^{2}\left( 2\left( g^{2}r^{2}\right) ^{\prime }+4rg^{2}\frac{%
f^{\prime }}{f}+r^{2}\left( g^{2}\right) ^{\prime }\frac{f^{\prime }}{f}%
+2r^{2}g^{2}\frac{f^{\prime \prime }}{f}\right)  \notag \\
& \qquad +\text{sgn}(\alpha )\,l^{2}r^{2}\left( 3\left( g^{4}\right)
^{\prime }\frac{f^{\prime }}{f}+4g^{4}\frac{f^{\prime \prime }}{f}\right)
\left( g^{4}\right) ^{\prime },  \label{24.5} \\
B_{2}(r)& =2r\left\{ 2-\text{sgn}(\alpha )\,l^{2}\left( \left( g^{2}\right)
^{\prime }\frac{f^{\prime }}{f}+2g^{2}\frac{f^{\prime \prime }}{f}\right)
\right\} .  \label{24.6}
\end{align}

The equation (\ref{23.4}) with $u=0$ can be rewritten as 
\begin{equation*}
-\frac{A_{0}(r)}{B_{0}(r)}=\frac{\tilde{R}(\tilde{x})}{6}.
\end{equation*}

Since the left side depends only on $r$ and the right side depends only on $%
\tilde{x}$, we have that both sides must be equal to a constant $\gamma $,
so that 
\begin{equation}
\tilde{R}(\tilde{x})=6\gamma .  \label{25}
\end{equation}

An Einstein manifold $\Sigma _{n}$ is a Riemannian or pseudo Riemannian
manifold whose Ricci tensor is proportional to the metric 
\begin{equation}
\tilde{R}_{\mu \nu }=kg_{\mu \nu }.  \label{25a}
\end{equation}%
The contraction of eq. (\ref{25a}) with the inverse metric $g^{\mu \nu }$
reveals that the constant of proportionality $k$ is related to the scalar
curvature $\tilde{R}$ by 
\begin{equation}
\tilde{R}=nk,  \label{25b}
\end{equation}%
where $n$ is the dimension of $\Sigma _{n}$.

Introducing (\ref{25a}) and (\ref{25b}) into the so called contracted
Bianchi identities, 
\begin{equation*}
\tilde{\partial}^{\beta }\left( \tilde{R}_{\alpha \beta }-\frac{1}{2}%
g_{\alpha \beta }\tilde{R}\right) =0,  \label{25c}
\end{equation*}%
we find 
\begin{equation*}
\left( n-2\right) \tilde{\partial}_{\beta }k=0.  \label{25d}
\end{equation*}%
This means that if $\Sigma _{n}$ is a Riemannian manifoldof dimension $n>2$
with metric $g_{\alpha \beta }$, then $k$ must be a constant.

On the other hand, in a $n$-dimensional space, the Riemann tensor can be
decomposed into its irreducible components 
\begin{eqnarray}
\tilde{R}_{\mu \nu \rho \sigma } &=&\tilde{C}_{\mu \nu \rho \sigma }+\frac{1%
}{n-2}\left( g_{\mu \rho }\tilde{R}_{\nu \sigma }-g_{\mu \sigma }\tilde{R}%
_{\nu \rho }-g_{\nu \rho }\tilde{R}_{\mu \sigma }+g_{\nu \sigma }\tilde{R}%
_{\mu \rho }\right)  \notag \\
&&+\frac{1}{\left( n-1\right) \left( n-2\right) }\left( g_{\mu \sigma
}g_{\nu \rho }-g_{\mu \rho }g_{\nu \sigma }\right) \tilde{R},  \label{25e}
\end{eqnarray}%
where $\tilde{C}_{\mu \nu \rho \sigma }$ is the Weyl conformal tensor, $%
\tilde{R}_{\alpha \beta }$ is the Ricci tensor and $\tilde{R}$ is the Ricci
scalar curvature.

Introducing (\ref{25a}), (\ref{25b}) into (\ref{25e}) we have 
\begin{equation}
\tilde{R}_{\mu \nu \rho \sigma }=\tilde{C}_{\mu \nu \rho \sigma }+\kappa
\left( g_{\mu \rho }g_{\nu \sigma }-g_{\mu \sigma }g_{\nu \rho }\right) ,
\label{25f}
\end{equation}%
where $\kappa =k/(n-1)$.

From (\ref{25f}) we can see that when $\tilde{C}_{\mu \nu \rho \sigma }=0$,
the Einstein manifold $\Sigma _{n}$ is a Riemannian manifold with constant
curvature $\kappa $.

Since the Weyl tensor is identically zero when $n=3$, we have that, if $n=3$%
, there is no distinction between Einstein manifolds and constant curvarture
manifolds. However, for $n>3$, constant curvature manifolds are special
cases of Einstein manifolds. This means that our $\Sigma _{3}(\tilde{x})$
manifold is a Riemannian manifold of constant curvature $\kappa =6\gamma$.

The solution of $A_{0}(r)/B_{0}(r)=-\gamma $ leads to 
\begin{equation*}
g^{2}(r)=\gamma +\text{sgn}(\alpha )\left( \frac{r^{2}}{l^{2}}-\beta \sqrt{%
\frac{r^{4}}{l^{4}}+\text{sgn}(\alpha )\frac{\mu }{l^{4}}}\right) ,
\end{equation*}%
where $\mu $ is a constant of integration and $\beta =\pm 1$. The equations (%
\ref{23.4}) with $u=0$ and $u=1$ lead to $f^{2}(r)=g^{2}(r)$, while $u=2$
tells us that 
\begin{equation*}
\tilde{R}=6\lambda ,
\end{equation*}%
where the constant of integration $\lambda $ must be equal to $\gamma $, so
that is consistent with eq. (\ref{25}).

In short, if the line element is given by (\ref{18}), then the functions $%
f(r)$ and $g(r)$ are given by 
\begin{equation}
f^{2}(r)=g^{2}(r)=\gamma +\text{sgn}(\alpha )\frac{r^{2}}{l^{2}}-\text{sgn}%
(\alpha )\,\beta \sqrt{\frac{r^{4}}{l^{4}}+\text{sgn}(\alpha )\frac{\mu }{%
l^{4}}}  \label{29}
\end{equation}%
where $\beta =\pm 1$ shows the degeneration due to the quadratic character
of the field equations, $\mu $ is a constant of integration related to the
mass of the system and $\gamma $ is another integration constant related to
the scalar curvature of the base manifold ($\tilde R=6\gamma$): $\gamma =0$
if it is flat, $\gamma =-1$ if it is hyperbolic (negative curvature) or $%
\gamma =1$ if it is spherical (positive curvature).

\subsection{A solution for equation (\protect\ref{3.2})}

Since the explicit form of the $h^{a}$ field is important in an eventual
construction of the matter lagrangian $L_{M}$, we are interested in to solve
the field equation (\ref{3.2}) for the $h^{a}$ field. 
\begin{equation}
\varepsilon _{abcde}R^{cd}\mathrm{D}_{\omega }h^{e}=0.  \label{eccampoh01}
\end{equation}

Expanding the $h^{a}=h_{\text{ }\mu }^{a}dx^{\mu }$ field in their holonomic
index, we have 
\begin{equation*}
h_{a}=h_{\mu \nu }e_{a}^{\mu }\,\mathrm{d}x^{\nu }.
\end{equation*}

For the space-time with a three-dimensional manifold maximally symmetrical $%
\Sigma _{3}$, we will assume that the field $h_{\mu \nu }$ must satisfy the
Killing equation $\mathcal{L}_{\xi }h_{\mu \nu }=0$ for $\xi _{0}=\partial
_{t}$ (stationary) and the six generators of the $\Sigma _{3}$, i.e., we are
assuming that the field $h_{\mu \nu }$ has the same symmetries than the
metric tensor $g_{\mu \nu }$.

\subsubsection{\textbf{Killing vectors of $\Sigma_3$ and shape of field $h^a$%
}}

When the curvature of $\Sigma _{3}$ is $\gamma =1$ (spherical type), it can
show that its \textit{driebein} is given by 
\begin{equation*}
\tilde{e}^{1}=\text{d}x_{1}\ ,\ \tilde{e}^{2}=\sin (x_{1})\,\text{d}x_{2}\
,\ \tilde{e}^{3}=\sin (x_{1})\sin (x_{2})\,\text{d}x_{3},  \label{dri01}
\end{equation*}
whose Killing vectors are \cite{salg4,salg6} 
\begin{align}
\xi _{1}& =\partial _{x_{3}},  \notag \\
\xi _{2}& =\sin x_{3}\,\partial _{x_{2}}+\cot x_{2}\cos x_{3}\,\partial
_{x_{3}},  \notag \\
\xi _{3}& =\sin x_{2}\sin x_{3}\,\partial _{x_{1}}+\cot x_{1}\cos x_{2}\sin
x_{3}\,\partial _{x_{2}}+\cot x_{1}\csc x_{2}\cos x_{3}\,\partial _{x_{3}},
\label{kv01} \notag\\
\xi _{4}& =\cos x_{3}\,\partial _{x_{2}}-\cot x_{2}\sin x_{3}\,\partial
_{x_{3}},  \notag \\
\xi _{5}& =\sin x_{2}\cos x_{3}\,\partial _{x_{1}}+\cot x_{1}\cos x_{2}\cos
x_{3}\,\partial _{x_{2}}-\cot x_{1}\csc x_{2}\sin x_{3}\,\partial _{x_{3}}, 
\notag \\
\xi _{6}& =\cos x_{2}\,\partial _{x_{1}}-\cot x_{1}\sin x_{2}\,\partial
_{x_{2}}.  \notag
\end{align}

On the other hand, when the curvature of $\Sigma_3$ is $\gamma=-1$
(hyperbolic type), its \textit{driebein} and their Killing vectors are the
same of the spheric type just changing the trigonometrical functions of $x_1$
for hyperbolical ones. For example, in this case $\tilde e^3=\sinh
(x_1)\sin(x_2)\,\text{d}x_3$.

The third case, $\gamma =0$ is the simplest. The \emph{driebein} is given by 
\begin{equation*}
\tilde{e}^{1}=\text{d}x_{1}\ ,\ \tilde{e}^{2}=\text{d}x_{2}\ ,\ \tilde{e}%
^{3}=\text{d}x_{3}  \label{dri02}
\end{equation*}
and their Killing vectors are given by 
\begin{align*}
\xi _{1}& =\partial _{x_{1}}\ ,\ \xi _{2}=\partial _{x_{2}}\ ,\ \xi
_{3}=\partial _{x_{3}},  \notag \\
\xi _{4}& =-x_{3}\,\partial _{x_{2}}+x_{2}\,\partial _{x_{3}},  
\\
\xi _{5}& =x_{3}\,\partial _{x_{1}}-x_{1}\,\partial _{x_{3}},  \notag \\
\xi _{6}& =-x_{2}\,\partial _{x_{1}}+x_{1}\,\partial _{x_{2}}.  \notag
\end{align*}

Then, we have 
\begin{align}
h^{T}& =h_{t}(r)\,e^{T}+h_{tr}(r)\,e^{R},  \notag \\
h^{R}& =h_{rt}(r)\,e^{T}+h_{r}(r)\,e^{R},  \label{hfield01} \\
h^{m}& =rh(r)\,\tilde{e}^{m}.  \notag
\end{align}

\subsubsection{\textbf{Dynamic of the field $h^a$}}

In order to obtain the dynamics of the field $h^{a}$ found in (\ref{hfield01}%
), we must replace this and the $2-$form curvature (\ref{22}) in the field
equation (\ref{eccampoh01}). Depending of the curvature of $\Sigma _{3}$ two
cases are possible.

First, if $\gamma =0$ the equation (\ref{eccampoh01}) is satisfied
identically. This means that the nonzero components of $h^{a}$ field given
in equation (\ref{hfield01}) are not determined by field equations. Second,
if $\gamma =\pm 1$, the equation (\ref{eccampoh01}) leads to the following
conditions 
\begin{align}
h_{tr}& =h_{rt}=0,  \label{C.9} \\
h_{r}& =(rh)^{\prime },  \label{C.10} \\
(fh_{t})^{\prime }& =f^{\prime }h_{r}.  \label{C.11}
\end{align}

From eq. (\ref{C.11}), we obtain 
\begin{equation*}
h_{t}(r)=h_{r}(r)-\frac{1}{f(r)}\int h_{r}^{\prime }(r)\,f(r)\,dr+\frac{A}{%
f(r)},
\end{equation*}%
where $A$ is a constant to be determined and we have performed integration
by parts. Then, we can solve equation (\ref{C.10}) 
\begin{equation*}
h(r)=\frac{1}{r}\int h_{r}(r)\,dr+\frac{B}{r},
\end{equation*}
where $B$ is another integration constant and $f(r)$ is the vielbein
component $e_{t}^{T}$.

Again, we realize that not all the nonzero components of $h^{a}$ field are
determined by the field equations.

The simplest case happens when $h_r$ is constant, namely $h_r(r)=h_0$. The
other components of $h^a$ field are 
\begin{equation*}
h_t(r)=h_0+\frac{A}{f(r)}\,,\quad h(r)=h_0+\frac{B}{r},
\end{equation*}
whose asymptotic behavior is given by 
\begin{equation*}
h_r(r\rightarrow\infty)=h_0\,,\quad h_t(r\rightarrow\infty)=h_0+\gamma
A\,,\quad h(r\rightarrow\infty)=h_0.
\end{equation*}

\subsection{Energy-momentum tensor for the field $h^{a}$}

From the vielbein found in the previous section we can find the
energy-momentum tensor associated to the field $h^{a}$, i.e., we can solve
the equation (\ref{7.3}). Let us suppose that the energy-momentum tensor
associated to the field $h^{a}$ can be modeled as an anisotropic fluid. In
this case, the components of the energy-momentum tensor can be written in
terms of the density of matter and the radial and tangential pressure. In
the frame of reference comoving, we obtain 
\begin{equation}
\hat{T}_{TT}^{(h)}=\rho ^{(h)}(r),\quad \hat{T}_{RR}^{(h)}=p_{R}^{(h)}(r),%
\quad \hat{T}_{ii}^{(h)}=p_{i}^{(h)}(r).  \label{45tmunu01}
\end{equation}

Considering these definitions with the solution found in (\ref{29}) and
replacing in the field equations (\ref{7.3}), we obtain 
\begin{align}
\rho ^{(h)}(r)& =-p_{R}^{(h)}(r)=-\frac{12}{l^{2}\kappa _{\text{EH}}}\left\{
2-\beta \frac{2+\text{sgn}(\alpha )\frac{\mu }{r^{4}}}{\sqrt{1+\text{sgn}%
(\alpha )\frac{\mu }{r^{4}}}}\right\} ,  \label{36.1} \\
p_{i}^{(h)}(r)& =\frac{4}{l^{2}\kappa _{\text{EH}}}\left\{ 6-\beta \,\frac{%
6+9\,\text{sgn}(\alpha )\frac{\mu }{r^{4}}+\frac{\mu ^{2}}{r^{8}}}{\left( 1+%
\text{sgn}(\alpha )\frac{\mu }{r^{4}}\right) ^{\frac{3}{2}}}\right\} .
\label{36.3}
\end{align}

Note that equations (\ref{36.1}) and (\ref{36.3}) show that the energy
density and the pressures do not depend on the $\gamma $ constant (See
appendix \ref{ape}).

\subsection{Energy density and radial pressure \label{edrp01}}

Now consider the conditions that must be satisfied by the energy density $%
\rho ^{(h)}(r)$ and radial pressure $p_{i}^{(h)}(r)$. From eq. (\ref{36.1})
we can see that the energy density is zero for all $r$, only if $\beta =1$
and $\mu =0$. This is the only one case where $\rho ^{(h)}(r)$ vanishes.
Otherwise the energy density is always greater than zero or always less than
zero.

In order to simplify the analysis, the energy density can be rewritten as 
\begin{equation}
\rho ^{(h)}(r) =-\frac{12}{l^{2}\kappa _{\text{EH}}}\left\{ \frac{2\sqrt{1+%
\text{sgn}(\alpha )\frac{\mu }{r^{4}}}-\beta \left( 2+\text{sgn}(\alpha )%
\frac{\mu }{r^{4}}\right) }{\sqrt{1+\text{sgn}(\alpha )\frac{\mu }{r^{4}}}}%
\right\} .  \label{38}
\end{equation}
Since the solution found in (\ref{29}) has to be real, then it must be
satisfied that $1+\text{sgn}(\alpha )\frac{\mu }{r^{4}}>0$. This implies
that the terms which appear in the numerator of eq. (\ref{38}) satisfy the
following constraint

\begin{equation*}
0<2\sqrt{1+\text{sgn}(\alpha )\frac{\mu }{r^{4}}}<\left( 2+\text{sgn}(\alpha
)\frac{\mu }{r^{4}}\right) .
\end{equation*}

This constraint is obtained by considering that $\left( \text{sgn}(\alpha )%
\frac{\mu }{r^{4}}\right) ^{2}>0$, adding to both sides $4\left( 1+\text{sgn}%
(\alpha )\frac{\mu }{r^{4}}\right) $ and then taking the square root. So, if 
$\beta =-1$ we can ensure that the energy density is less than zero. If $%
\beta =1$ the energy density is greater than zero, unless that $\mu =0$,
case in that the energy density is zero. The radial pressure behaves exactly
reversed as was found in eq. (\ref{36.1}).

We also can see if $\mu =0$ the energy density remains constant. Otherwise,
the energy density is a monotonic increasing ($\beta =-1$) or decreasing ($%
\beta =1$) function of radial coordinate.

Note that if $\beta =-1$ then when $r\rightarrow \infty ,$ the energy
density and the radial pressure tend a nonzero value 
\begin{equation*}
\rho ^{(h)}(r\rightarrow \infty )=-p_{R}^{(h)}(r\rightarrow \infty )=-\frac{%
48}{l^{2}\kappa _{\text{EH}}},
\end{equation*}
as if it were a negative cosmological constant. Otherwise, $\beta =+1$, the
energy density and the radial pressure are asymptotically zero, as in the
case of a null cosmological constant.

\begin{figure}[h]
\includegraphics[width=0.7\columnwidth]{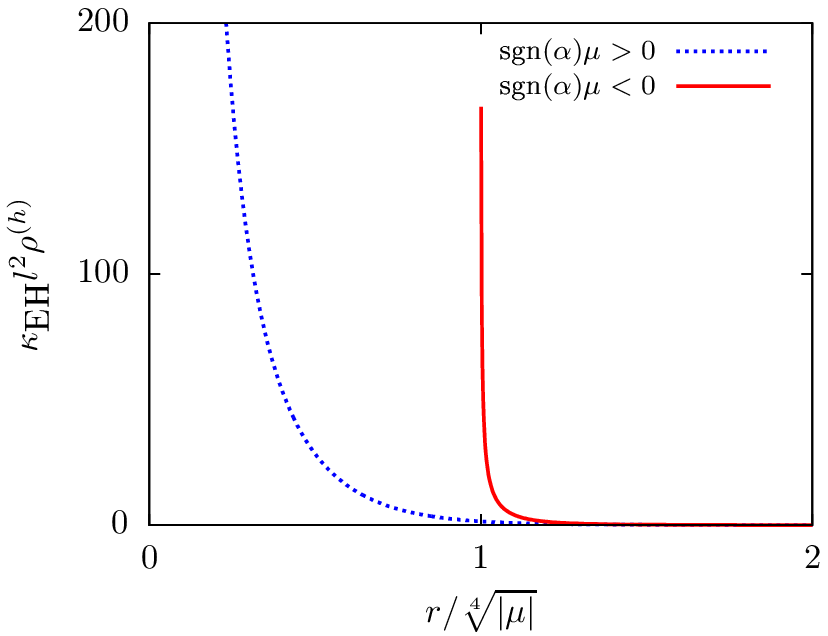} \centering
\caption{The energy density associated with the field $h^a$ ($\protect\beta%
=1 $).}
\label{temfig01a}
\end{figure}

\begin{figure}[h]
\includegraphics[width=0.7\columnwidth]{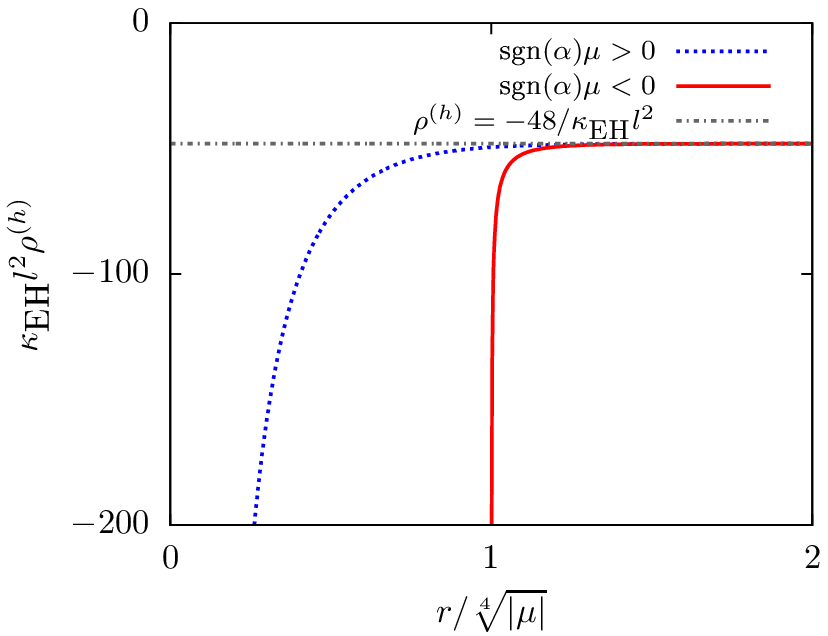} \centering
\caption{The energy density associated with the field $h^a$ ($\protect\beta%
=-1$).}
\label{temfig01b}
\end{figure}

In summary,

\begin{itemize}
\item If $\mu =0$, then the energy density is constant throughout the space,
zero if $\beta=1$ and $-\frac{48}{l^{2}\kappa _{\text{EH}}}$ if $\beta=-1$.

\item If $\beta =1$ and $\mu\neq 0$, the energy density is positive and
decreases to zero at infinity (see figure \ref{temfig01a}).

\item If $\beta =-1$ and $\mu\neq 0$, the energy density is negative and its
value grows to $-\frac{48}{l^{2}\kappa _{\text{EH}}}$ (see figure \ref{temfig01b}).
\end{itemize}

As we have already shown, the radial pressure is the negative of energy
density.

\subsection{Tangential pressures}

\label{tp01}

We can see that the tangential pressures given in the eq. (\ref{36.3})
vanishes if 
\begin{equation*}
\text{sgn}(\alpha )\frac{\mu }{r^{4}}=9+4\beta \sqrt{6}.
\end{equation*}

Thus we have

\begin{figure}[h]
\includegraphics[width=0.7\columnwidth]{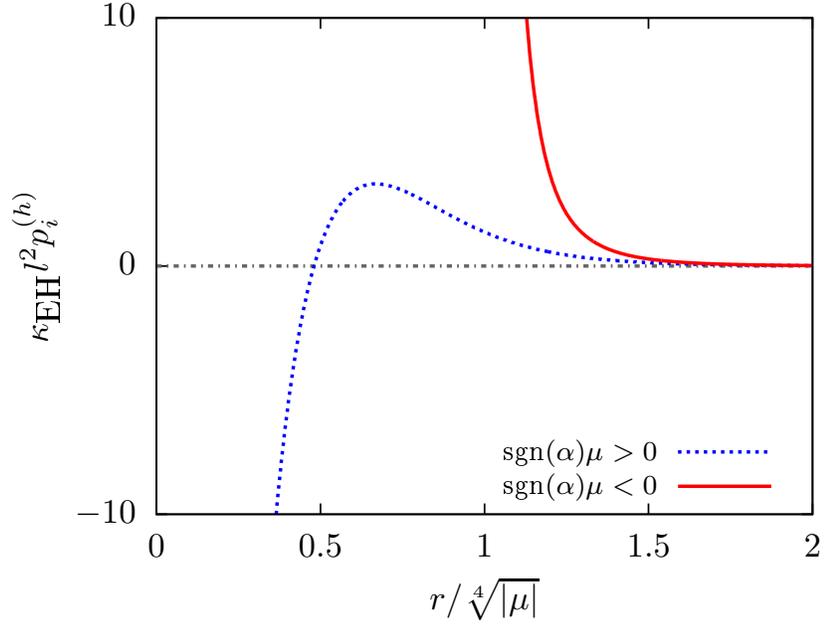} \centering
\caption{The tangential pressures associated with the field $h^a$ ($\protect%
\beta=1$).}
\label{temfig03a}
\end{figure}

\begin{figure}[h]
\includegraphics[width=0.7\columnwidth]{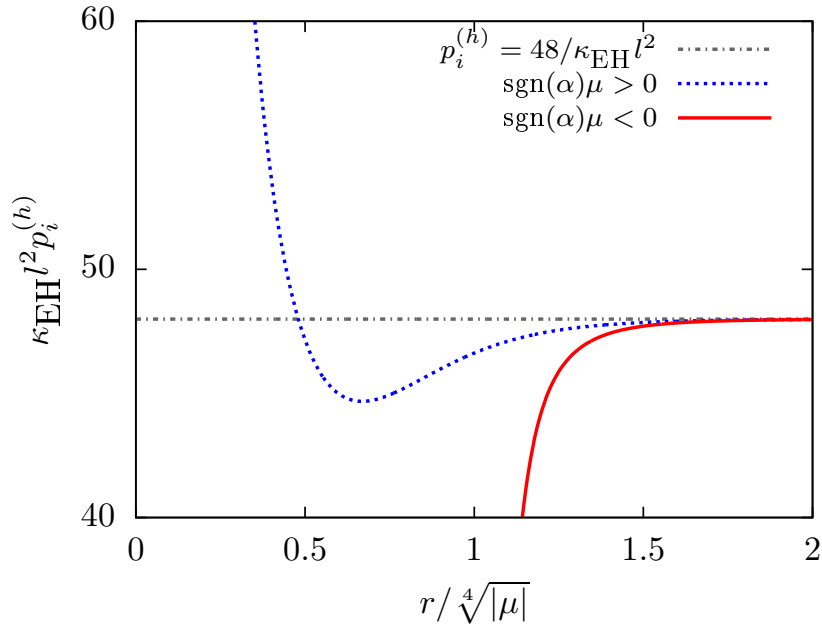} \centering
\caption{The tangential pressures associated with the field $h^a$ ($\protect%
\beta=-1$).}
\label{temfig03b}
\end{figure}

\begin{itemize}
\item If $\beta =1,$ the tangential pressure vanishes only if $\text{sgn}%
(\alpha )\mu $ is greater than zero (see figure \ref{temfig03a}).

\item If $\beta =-1$ the tangential pressure vanishes only if $\text{sgn}%
(\alpha )\mu $ is less than zero (see figure \ref{temfig03b}).

\item In other cases, the tangential pressure does not change sign.
\end{itemize}

Furthermore, it is straightforward to show that there is only one critical
point at $r=\sqrt[4]{\frac{\text{sgn}(\alpha )\mu }{5}}$ only if $\mathrm{sgn%
}(\alpha )\mu >0$.

\subsubsection{\textbf{Case $\protect\beta =1$}}

If \textbf{$\beta =1,$} three cases are distinguished, depending on the
quantity $\mathrm{sgn}(\alpha )\mu $

\begin{enumerate}[($a$)]
\item For $\mu =0$, we have the simplest case. The tangential pressure
is zero for all $r$ .

\item If $\mathrm{sgn}(\alpha )\mu >0,$ the tangential pressure
diverges at $r=0$. \ It is a function that tends to $-\infty $ at $r=0$,
vanishes at 
\begin{equation}
r_{1}=\sqrt[4]{\text{sgn}(\alpha )\mu \,\frac{4\sqrt{6}-9}{15}}\approx 0.48%
\sqrt[4]{|\mu |},  \label{r101}
\end{equation}
takes its maximum value 
\begin{equation*}
p_{i}^{(h)\,\text{max}}=\frac{4}{9\ l^{2}\kappa _{\text{EH}}}\left( 54-19%
\sqrt{6}\right) \approx \frac{3.3}{\ l^{2}\kappa _{\text{EH}}}
\end{equation*}%
at 
\begin{equation}
r_{2}=\sqrt[4]{\frac{\text{sgn}(\alpha )\mu }{5}}\approx 0.67\sqrt[4]{|\mu |}
\label{r201}
\end{equation}%
and decreases to zero when $r$ tends to infinity.

\item If $\mathrm{sgn}(\alpha )\mu <0$, then the tangential pressure
tends to $+\infty $ at 
\begin{equation*}
r_{m}=\sqrt[4]{-\text{sgn}(\alpha )\mu }=\sqrt[4]{|\mu |}.
\end{equation*}%
Of course, the manifold is not defined for $r<r_{m}$ (see the metric
coefficients in eq. (\ref{29})). The tangential pressure is a decreasing
function of $r$ which vanishes at infinity, but always greater than zero.
\end{enumerate}

\subsubsection{\textbf{Case $\protect\beta =-1$}}

If \textbf{$\beta =-1$} three situations are also distinguished

\begin{enumerate}[($a$)]
\item For $\mu =0$, we have the simplest case. The tangential pressure
is constant and greater than zero for all $r$ 
\begin{equation*}
p_{i}^{(h)}(r)=\frac{48}{l^{2}\kappa _{\text{EH}}}.
\end{equation*}

\item If $\mathrm{sgn}(\alpha )\mu >0$, the tangential pressure
diverges to positive infinity at $r=0,$ is a decreasing function of $r$,
reaches a minimum value 
\begin{equation*}
p_{i}^{(h)\,\text{min}}=\frac{4}{9\ l^{2}\kappa _{\text{EH}}}\left( 54+19%
\sqrt{6}\right) \approx \frac{45}{\ l^{2}\kappa _{\text{EH}}}
\end{equation*}%
at 
\begin{equation*}
r=\sqrt[4]{\frac{\text{sgn}(\alpha )\mu }{5}}\approx 0.67\sqrt[4]{|\mu |},
\end{equation*}%
and then increases to a bounded infinite value 
\begin{equation*}
p_{i}^{(h)}(r\rightarrow \infty )=\frac{48}{l^{2}\kappa _{\text{EH}}}.
\end{equation*}%
The tangential pressure is always greater than zero.

\item If $\mathrm{sgn}(\alpha )\mu <0,$ the tangential pressure
diverges to negative infinity at (remember that the manifold is not defined
for $r<r_{m}$) 
\begin{equation*}
r_{m}=\sqrt[4]{-\text{sgn}(\alpha )\mu }=\sqrt[4]{|\mu |}.
\end{equation*}%
The tangential pressure is an increasing function of $r$ which tends to a
positive constant value when $r$ goes to infinity 
\begin{equation*}
p_{i}^{(h)}(r\rightarrow \infty )=\frac{48}{l^{2}\kappa _{\text{EH}}}.
\end{equation*}

Furthermore, the tangential pressures become zero at 
\begin{equation*}
\quad r=\sqrt[4]{-\text{sgn}(\alpha )\mu \,\frac{9+4\sqrt{6}}{15}}\approx
1.06\sqrt[4]{|\mu |}.
\end{equation*}
\end{enumerate}

\section{Spherically Symmetric Solution from General Solution \label%
{sectionIV}}

Now consider the case of spherically symmetric solutions studied in Ref. 
\cite{salg4} and reviewed in section \ref{sectionII}. These solutions are
described by the vielbein defined in eq. (\ref{10}) with the functions $f(r)$
and $g(r)$ given in eq. (\ref{14}).

This solution corresponds to the general static solution found in (\ref{29})
where
\begin{inparaenum}[(i)]
	\item  the curvature of the so called, three-dimensional base manifold,
	is taken positive $\gamma =1$ (sphere $S^{3}$),
	\item  the constant $\mu $,
	written in terms of the mass $M$ of the distribution is given by 
	\begin{equation*}
	\mu =\frac{\kappa _{EH}}{6\pi ^{2}}Ml^{2}>0,
	\end{equation*}
	\item  and $\beta =1$ so that this solution has as limit when $%
	l\rightarrow 0$, the 5D Schwarzschild black hole obtained from the Einstein
	Hilbert gravity.
\end{inparaenum}

From Ref. \cite{salg4}, we know that the relative values of the mass $M$ and
the distance $l$ of this solution leads to black holes or naked
singularities.

\begin{enumerate}[($a$)]
\item In the event that $\alpha >0$, the manifold only has one
singularity at $r=0$. Otherwise, if $\alpha <0$, the manifold has only one
singularity at 
\begin{equation}\label{rm01}
r_{m}=\sqrt[4]{\mu }=\sqrt[4]{\frac{\kappa _{\mathrm{EH}}}{6\pi ^{2}}Ml^{2}}.
\end{equation}

\item There is a black hole solution with event horizon defined by 
\begin{equation}
r_{0}=\sqrt{\frac{\mu -\text{sgn}(\alpha )\,l^{4}}{2l^{2}}}=\sqrt{\frac{%
\kappa _{\text{EH}}}{12\pi ^{2}}M-\text{sgn}(\alpha )\frac{l^{2}}{2}},
\label{bh01}
\end{equation}%
if $\mu >l^{4}$, or equivalently 
\begin{equation}
\frac{\kappa _{\mathrm{EH}}}{6\pi ^{2}}M>l^{2}.  \label{bh02}
\end{equation}
Otherwise, there is a naked singularity.
\end{enumerate}

\subsection{Case $\protect\alpha >0$}

In this case the energy density appears to be decreasing and vanishes at
infinity and the radial pressure behaves reversed (see subsection \ref{edrp01} with $\beta=1$ and $\mathrm{sgn}(\alpha)\,\mu>0$).

Much more interesting is the behavior of the tangential pressure. In fact,
as we already studied in subsection \ref{tp01}, the tangential pressure is
less than zero for $r<r_{1}$ (\ref{r101}), vanishes at $r_1$, becomes
greater than zero until reaching a maximum at $r_{2}$ (\ref{r201}) and then
decreases until it becomes zero at infinity.

\subsubsection{\textbf{Comparison between $r_{0}$, $r_{1}$ and $r_{2}$ for
black hole solution}}

When the solution found is a black hole, then it must satisfy the condition (%
\ref{bh02}) and has event horizon in $r_{0}$ given in (\ref{bh01}). It may
be of interest to study the cases when $r_{0}$ is larger or smaller than $%
r_{1}$ and $r_{2}$.

First consider $r_{0}$ for $l$ fixed, i.e., we study the behavior of the $%
r_{0}=r_{0}(\mu )$ function. For $\mu \geq l^{4}$ (black hole solution), $%
r_{0}=r_{0}(\mu )$ is a well-defined, continuous and strictly increasing
function of $\mu$ which has an absolute minimum at $\mu =l^{4}$, where it
vanishes, i.e., $r_{0}(\mu =l^{4})=0$. Furthermore, when $\mu \gg l^{4}$ the 
$r_0(\mu)$ function behaves like $\sqrt{\mu }$.

On the other hand, the study of functions $r_{1}(\mu )$ and $r_{2}(\mu )$
shows that they are well defined, continuous and strictly increasing
functions of $\mu\geq 0 $ which vanish at $\mu =0$. As $\mu$ increases, $%
r_{1}$ and $r_{2}$ grow proportional to $\sqrt[4]{\mu }$.

From the definitions of $r_1$ and $r_2$ given in eqs. (\ref{r101}) and (\ref%
{r201}), and the preceding analysis, it follows that $r_{2}>r_{1}>r_{0}$ if $%
\mu =l^{4}$, and $r_{0}>r_{2}>r_{1}$ if $\mu \rightarrow \infty $. This
means that should exist a unique value of the constant $\mu $, denoted $\mu
_{1}$ such that $r_{0}(\mu _{1})=r_{1}(\mu _{1})$ and a single $\mu _{2}$
such that $r_{0}(\mu _{2})=r_{2}(\mu _{2})$. After some calculations is
obtained 
\begin{equation*}
\mu _{1}=\frac{l^{4}}{15}\left( 8\sqrt{6}-3+2\sqrt{6\left( 7-2\sqrt{6}%
\right) }\right) \approx 1.58\ l^{4}
\end{equation*}%
and 
\begin{equation*}
\mu _{2}=\frac{l^{4}}{5}\left( 7+2\sqrt{6}\right) \approx 2.38\ l^{4}.
\end{equation*}

From the above analysis it is concluded that depending on the value of the
constant $\mu $, proportional to the mass, we could have the following cases

\begin{itemize}
\item If $l^{4}\leq \mu <\mu _{1}$ then $r_{0}<r_{1}$. Outside the black
hole horizon, there is a region $r_{0}<r<r_{1}$ where the tangential
pressure is negative.

\item If $\mu >\mu _{1}$ then $r_{0}>r_{1}$, the zone in which the
tangential pressure is negative is enclosed within the black hole horizon.
\end{itemize}

A completely analogous analysis can be done to study the relationship
between $r_{0}$ and $r_{2}$: if $\mu<\mu_2$, the maximum value of the
tangential pressure is outside the event horizon or, inside if $\mu>\mu_2$.

\subsubsection{\textbf{Pressure radial and tangential pressures }}

\begin{figure}[h]
\centering
\includegraphics[width=0.7\columnwidth]{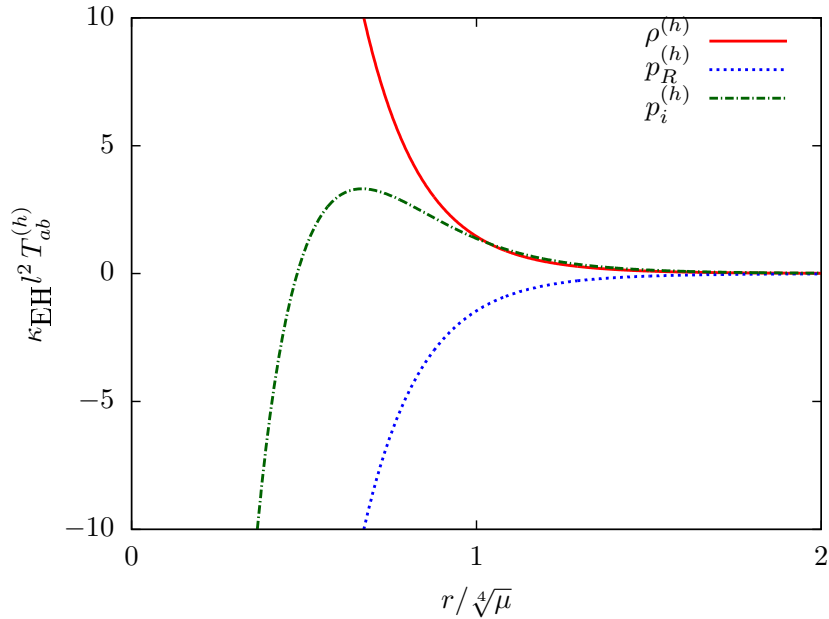}
\caption{The components of energy-momentum tensor associated with the field $%
h^a$ ($\protect\alpha >0$). Note the interesting behavior of tangential
pressure: is negative for $r<r_1$, vanishes at $r_1$, reaches a maximum at $%
r_2$ and then decreases to zero at infinity. The energy density and
pressures tend rapidly to zero as $1/r^{8}$.}
\label{amc01}
\end{figure}

In summary, for $\alpha>0$ we can see that the energy density is always
greater than zero, while the radial pressure is less than zero, both vanish
when $r$ goes to infinity (see figure \ref{amc01}).

On the other hand, the lateral pressures are less than zero for $r<r_1$,
become positive for $r>r_1$ reaching a maximum at $r_2$ and then decrease
until vanish when $r$ goes to infinity (see figure \ref{amc01}).

The solution may be a naked singularity $(\mu <l^{4})$ or a black hole $(\mu
>l^{4})$. In case of a black hole there is an event horizon at $r=r_{0}$,
which can hide the zone of negative tangential pressures ($\mu >\mu _{1}$)
or otherwise, remains uncovered.

\subsection{Case $\protect\alpha <0$}

Now consider the coupling constant $\alpha <0$. In this case the space-time
has a minimum radius $r_m$, defined in (\ref{rm01}), where is located the
singularity.

\begin{figure}[h]
\centering
\includegraphics[width=0.7\columnwidth]{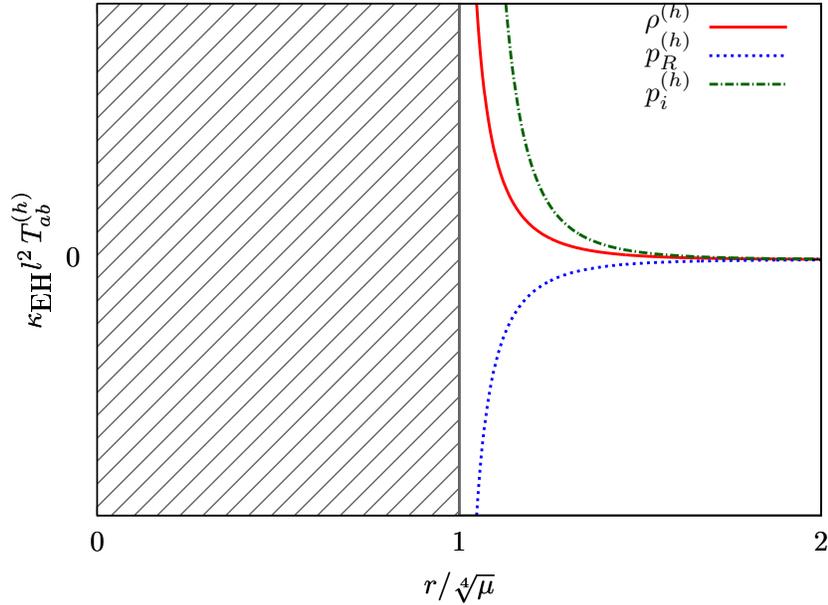}
\caption{The components of energy-momentum tensor associated with the field $%
h^a$ ($\protect\alpha <0$). The space time singularity is located at $r_m=%
\sqrt[4]{\protect\mu}$. The region $r <r_m$ does not belong to the variety.
As in case $\protect\alpha>0$ (see figure \ref{amc01}), the energy density
and pressures tend to zero as $1/r^8$.\label{fig06}}
\end{figure}

From analysis done in subsection \ref{edrp01} (with $\beta=1$ and $\mathrm{%
sgn}(\alpha)\,\mu<0$) we obtain that the energy density is progressively
reduced and vanishes at infinity. On the other hand, the radial pressure is
just the negative energy density (see figure \ref{fig06}).

Furthermore, the tangential positive pressure tends to infinity at $r=r_{m}=%
\sqrt[4]{\mu }$ and decreases to zero at infinity (see subsection \ref{tp01}
with $\beta=1$ and $\mathrm{sgn}(\alpha)\,\mu<0$).

\section{Concluding Remarks\label{sectionV}}

An interesting result of this work is that when the field $h^{a}$, which
appears in the Lagrangian (\ref{1}), is modeled as an anisotropic fluid (see
eqs. (\ref{36.1} - \ref{36.3})), we find that the solutions of the fields
equations predicts the existence of a negative tangential pressure zone
around low-mass distributions ($\mu <\mu _{1}$) when the coupling constant $%
\alpha $ is greater than zero.

Additionally ($\alpha >0$), this model predicts the existence of a maximum
in the tangential pressure, which can be observed in the outer region of a
field distribution that satisfies $\mu<\mu_{2}$.

It is also important to note that this model contains in its solutions
space, solutions that behave like those obtained from models with negative
cosmological constant ($\beta =-1$). In such a situation, the field $h^{a}$
is playing the role of a cosmological constant \cite{GMS,salg5}.

In this article we have assumed that the matter Lagrangian $L_{M}$ satisfies
the property $\delta L_{M}/\delta e^{a}=0$ and that the energy-momentum
tensor associated to the $h^{a}$ field can be modeled as an anisotropic
fluid. A possible explicit example of the matter Lagrangian which satisfies
these considerations could be constructed from a Lagrangian of the
electromagnetic field in matter \cite{11,12,13}. In fact, a
candidate for matter Lagrangian which satisfies the above conditions would be 
\begin{equation*}
L_{M}=-\frac{1}{2}\,\frac{1}{4!}\,\epsilon
_{cdmnl}F_{ab}H^{cd}h^{a}h^{b}h^{m}h^{n}h^{l},  
\end{equation*}%
where $F_{ab}$ is the electromagnetic field and $H^{cd}$ is the so called
electromagnetic exitation, which is given by (in tensorial notation) 
\begin{equation*}
H^{\mu \nu }=\frac{1}{2}\chi ^{\mu \nu \rho \sigma }F_{\rho \sigma },
\end{equation*}%
where the tensor density $\chi ^{\mu \nu \rho \sigma }$ describes the
electric and magnetic properties of matter.

\begin{acknowledgments}
This work was supported in part by FONDECYT Grants 1130653. Two of the
authors (F.G., C.Q.) were supported by grants from the Comisi\'{o}n Nacional
de Investiga\-ci\'{o}n Cien\-t\'{\i}\-fica y Tecnol\'{o}gica CONICYT and
from the Universidad de Concepci\'{o}n, Chile. One of the authors (P.M) was
supported by FONDECYT Grants 3130444.
\end{acknowledgments}

\appendix

\section{Obtaining the energy-momentum tensor associated to the field $h^a$%
\label{ape}}

In this appendix we show explicitly the computations that lead to obtain the
energy density and pressures associated to the field $h^a$ shown in eqs. (%
\ref{36.1}) and (\ref{36.3}) from the field equation (\ref{7.3}).

The field equation (\ref{7.3}) can be rewritten as 
\begin{equation}
-\frac{l^{2}}{2\kappa _{\text{EH}}}\text{sgn}(\alpha )\star \left( \epsilon
_{abcde}R^{bc}R^{de}\right) =\hat{T}_{ab}^{(h)}e^{b},  \label{eq-01}
\end{equation}%
where we see it is necessary to compute the components of the left side.
Straightforward calculations lead to 
\begin{align}
\star \left( \epsilon _{Tbcde}R^{bc}R^{de}\right) & =6\ \frac{\left\{
(g^{2}-\gamma )^{2}\right\} ^{\prime }}{r^{3}}\ e^{T},  \label{eq0601-01} \\
\star \left( \epsilon _{Rbcde}R^{bc}R^{de}\right) & =-6\ \frac{\left\{
(g^{2}-\gamma )^{2}\right\} ^{\prime }}{r^{3}}\ e^{R},  \label{eq0602-01} \\
\star \left( \epsilon _{ibcde}R^{bc}R^{de}\right) & =-2\ \frac{\left\{
(g^{2}-\gamma )^{2}\right\} ^{\prime \prime }}{r^{2}}\ e^{i}.
\label{eq0603-01}
\end{align}

From the general solution found in eq. (\ref{29}) we obtain 
\begin{equation}
(g^{2}-\gamma )^{2}=\frac{r^{4}}{l^{4}}\left( 2+\text{sgn}(\alpha )\frac{\mu 
}{r^{4}}-2\beta \sqrt{1+\text{sgn}(\alpha )\frac{\mu }{r^{4}}}\right) ,
\end{equation}%
so that 
\begin{equation}
\bigl\{(g^{2}-\gamma )^{2}\bigr\}^{\prime }=4\frac{r^{3}}{l^{4}}\left\{
2-\beta \frac{2+\text{sgn}(\alpha )\frac{\mu }{r^{4}}}{\sqrt{1+\text{sgn}%
(\alpha )\frac{\mu }{r^{4}}}}\right\}  \label{eq0604-01}
\end{equation}%
and 
\begin{equation}
\bigl\{(g^{2}-\gamma )^{2}\bigr\}^{\prime \prime }=4\frac{r^{2}}{l^{4}}%
\left\{ 6-\beta \,\frac{6+9\,\text{sgn}(\alpha )\frac{\mu }{r^{4}}+\frac{\mu
^{2}}{r^{8}}}{\left( 1+\text{sgn}(\alpha )\frac{\mu }{r^{4}}\right) ^{\frac{3%
}{2}}}\right\} ,  \label{eq0605-01}
\end{equation}%
where we realize that none of those last three terms depend on the $\gamma $
constant.

Hence, by replacing eqs. (\ref{eq0604-01}) and (\ref{eq0605-01}) into eqs. (%
\ref{eq0601-01} - \ref{eq0603-01}) and then into eq. (\ref{eq-01}), we get 
\begin{align*}
\rho ^{(h)}(r)& =-p_{R}^{(h)}(r)=-\frac{12}{l^{2}\kappa _{\text{EH}}}\left\{
2-\beta \frac{2+\text{sgn}(\alpha )\frac{\mu }{r^{4}}}{\sqrt{1+\text{sgn}%
(\alpha )\frac{\mu }{r^{4}}}}\right\} , \\
p_{i}^{(h)}(r)& =\frac{4}{l^{2}\kappa _{\text{EH}}}\left\{ 6-\beta \,\frac{%
6+9\,\text{sgn}(\alpha )\frac{\mu }{r^{4}}+\frac{\mu ^{2}}{r^{8}}}{\left( 1+%
\text{sgn}(\alpha )\frac{\mu }{r^{4}}\right) ^{\frac{3}{2}}}\right\} ,
\end{align*}%
where we have introduced the energy-momentum tensor associated to the field $%
h^{a}$ given in eq. (\ref{45tmunu01}).

\end{document}